\documentclass[prl,twocolumn,amssymb,amsmath,floatfix]{revtex4-1}
\usepackage{graphicx}
\usepackage{bm}

\begin{document}

\title{Electrodynamic Response and Stability of Molecular Crystals}
\author{Bohdan Schatschneider$^1$}
\author{Jian-Jie Liang$^2$}
\author{Anthony M. Reilly$^3$}
\author{Noa Marom$^4$}
\author{Guo-Xu Zhang$^3$}
\author{Alexandre Tkatchenko$^3$}
\email{tkatchenko@fhi-berlin.mpg.de}
\affiliation{$^1$The Pennsylvania State University, Fayette-The Eberly Campus, USA \\
$^2$Accelrys Inc., 10188 Telesis Court, Suite 100 San Diego, CA 92121 USA \\
$^3$Fritz-Haber-Institut der Max-Planck-Gesellschaft, Faradayweg 4-6, 14195 Berlin, Germany \\
$^4$Institute for Computational Engineering and Sciences, The University of Texas at Austin, Austin, TX 78712, USA}

\begin{abstract}
We show that electrodynamic dipolar interactions, responsible for long-range fluctuations
in matter, play a significant role in the stability of molecular crystals.
Density functional theory calculations with van der Waals interactions determined 
from a semilocal ``atom-in-a-molecule'' model result in a large overestimation of the
dielectric constants and sublimation enthalpies for polyacene crystals from naphthalene to pentacene, 
whereas an accurate treatment of non-local electrodynamic response leads to an agreement with the measured values
for both quantities. Our findings suggest that collective response effects play a substantial
role not only for optical excitations, but also for cohesive properties of non-covalently bound
molecular crystals.
\end{abstract}

\maketitle

Polyacene molecular crystals form a fundamental class of aromatic solids,
and have been extensively studied as potential materials for
organic electronics~\cite{Forrest-Nature,ChemMater-review,LK-JF-PRB-2012}. 
It is understood that the optical properties of polyacenes are very
sensitive to long-range intra- and intermolecular electrodynamic interactions. 
This is reflected by shifts in the optical absorption frequencies upon increasing the molecule size or 
upon solid formation~\cite{OrgCrys-book}, and is further exhibited 
by the visible color of oligoacene crystals, which changes from transparent in naphthalene and anthracene, 
to bright orange in tetracene, and deep blue in pentacene~\cite{OrgCrys-book,JEAnthony-Angewandte-2007}. 
The optical absorption spectrum is directly related to the polarizability through 
the Kramers-Kronig transformation~\cite{Toll-PR-1956}. Therefore, the observed changes in the optical spectrum
upon crystallization of polyacenes are accompanied by a change in the molecular polarizability.
In addition, these changes in polarization should directly impact the crystal lattice energy. 
However, the effect of electrodynamic intermolecular interactions on the cohesive properties of molecular crystals 
remains poorly understood.
In this Letter, we show that the dipolar electrodynamic coupling
between polyacene molecules reduces the solid dielectric constant by 15\%,
and has an impact of up to 0.5 eV per molecule on the computed van der Waals energies and sublimation enthalpies
of these molecular crystals. Our results imply that
electrodynamic response is crucial for describing both the 
cohesive energy and the optical properties of molecular crystals. 

Polyacene crystals are extended aromatic networks characterized by polarizable $\pi$ clouds. 
Therefore, an appreciable part of the crystal lattice energy stems from ubiquitous attractive
vdW dispersion interactions. When studying the cohesion of molecular systems, for example using
density-functional theory (DFT)~\cite{Neumann,CSP2010} or classical potentials~\cite{Beran-review}, 
the vdW energy is typically computed using effective polarizabilities for hybridized ``atoms'' inside a molecule. 
It is common to approximate the frequency-dependent polarizability of every
atom using a single effective excitation frequency (also called the Uns{\"o}ld approximation~\cite{Stone-book}).
In this model, the dipole polarizability for atom $p$ is written as
\begin{equation}
\label{eqPade}
\alpha_p(i\omega) = \frac{\alpha_p[n(\mathbf{r})]}{1+(\omega/\omega_p[n(\mathbf{r})])^2} ,
\end{equation}
where $\alpha_p[n(\mathbf{r})]$ is the static polarizability of an atom $p$, and $\omega_p[n(\mathbf{r})]$ 
is the corresponding characteristic excitation frequency. In this equation we emphasize that the
effective parameters can be defined as functionals of the self-consistent electron density $n(\mathbf{r})$ as done
in the Tkatchenko-Scheffler (TS) method~\cite{TS-vdW}. Regardless of whether one treats $\alpha_p$ and $\omega_p$ as empirical
parameters or obtains them from the electron density, their values for different carbon atoms in 
polyacene molecules and crystals turn out to be essentially degenerate (the same holds for the
hydrogen atoms). This finding can be attributed to the rather similar local hybridization environment
that every atom ``feels'' inside polyacene molecules. 
This simplified model for the polarizability would lead to a similar optical absorption spectrum for different polyacenes,
in stark disagreement with experimental measurements and explicit
excited-state calculations~\cite{OrgCrys-book,JEAnthony-Angewandte-2007,pentacene-BSE}. 

The semilocal approximation for the polarizability in Eq.~(\ref{eqPade}) neglects the dynamic electric fields that an
atom experiences from all the other 
atoms inside a molecule or a crystal. Recently, an efficient parameter-free method was developed 
to include these \emph{screening} effects on the polarizability for non-metallic molecules and solids~\cite{MBD}. 
We model the environment as a dipole field and solve the resulting 
classical Dyson-like self-consistent screening (SCS) equation~\cite{Felderhof,Oxtoby-Gelbart,Thole}
\begin{eqnarray}
\label{eqScr}
\alpha^{\rm{SCS}}({\bf r}; i\omega) &=& \alpha^{\rm{TS}}({\bf r}; i\omega) + \alpha^{\rm{TS}}({\bf r}; i\omega) \nonumber \\
&\times& \int{d {\bf r'} \mathcal{T}({\bf r}-{\bf r'}) \alpha^{\rm{SCS}}({\bf r'}; i\omega) }
\end{eqnarray}  
where $\alpha^{\rm{TS}}({\bf r}; i\omega)$ is the sum of the TS effective atomic polarizabilities~\cite{TS-vdW}, 
and $\mathcal{T}({\bf r}-{\bf r'})$ is the dipole--dipole interaction tensor (Hartree atomic units are used throughout). 
Equation~(\ref{eqScr}) is discretized using atomic positions as a basis, and then solved directly and exactly
by inverting the tensor corresponding to the coupled dipoles modeled as quantum harmonic oscillators (QHO).
The QHO parameters are defined using the TS polarizability~\cite{MBD}.
The solution of Eq.~(\ref{eqScr}) yields the non-local molecular polarizability
tensor $\alpha_{pq,ij}^{\rm{SCS}}(i\omega)$, where indices $p$ and $q$ label the atoms while
indices $i$ and $j$ label the atomic Cartesian coordinates. The contraction of the molecular tensor for every atom $p$ yields the atomic polarizability
tensors $\alpha_{p,ij}^{\rm{SCS}}(i\omega)$. These tensors now include both the short-range hybridization effects from the TS method
and the long-range response screening from the solution of the SCS equation.

The electrodynamic response included upon solving the SCS equation (Eq.~\ref{eqScr}) allows one to correctly
capture two important contributions to the polarizability: (i) (de)polarization, and (ii) polarizability anisotropy.
The local ``atom-in-a-molecule'' polarizability as defined in Eq.~(\ref{eqPade}) leads to an essentially 
isotropic response for molecules and solids~\cite{MBD}. 
The directionality of the polarization, well known for polyacenes from experiments~\cite{pentacene-dielectric-anisotropy} and calculations~\cite{pentacene-BSE}, 
emerges from the intrinsic anisotropy of the molecular orbitals and the electrodynamic coupling between them. In the SCS formalism of
Eq.~(\ref{eqScr}), the anisotropy of the molecular polarizability stems from the coupling between fluctuating 
QHOs. For a set of small organic molecules, the SCS calculation significantly reduces the error in the molecular anisotropy to 23\% 
from 80\% in the TS method~\cite{MBD}. The emergence of polarizability anisotropy is the main 
effect brought by the inclusion of electrodynamic response effects for small gas-phase molecules. In the solid state or for
larger molecules, the situation is more complex.
In a crystalline environment every atom experiences the electric field from the atoms within the same molecule 
(similar to the gas phase), as well as the field produced by neighboring molecules. 
While the screening of the molecule in the crystal leads to an anisotropic polarizability,
one also typically finds an appreciable change in the \emph{isotropic} polarizability of the molecule
when compared to the gas phase. 
%Therefore, the response screening is expected to have an impact on the crystal stability as well, 
%since vdW dispersion interactions arise from interacting polarizabilities.

In order to assess the relative importance of electrodynamic response on the properties 
of non-covalently bound molecular crystals, we have chosen to examine a series of polyacene crystals, ranging from
naphthalene to pentacene. This choice allows us to study the evolution of response properties of
molecular crystals and their stability as a function of molecular size and crystal environment. 
Initial crystal structures for each polyacene were obtained from the lowest temperature 
data sets available in the Cambridge Structural Database~\cite{CSD}. 
The low-temperature polymorphs were chosen for tetracene and pentacene. 
The crystal unit cells and the internal geometries were fully optimized using 
DFT with the generalized gradient approximation of Perdew, Burke, and Ernzerhof (PBE)~\cite{PBE}
with vdW interactions treated using the TS method~\cite{TS-vdW} (denoted as PBE+vdW). 
The CASTEP code was used for all calculations~\cite{CASTEP,Erik}. Norm-conserving pseudopotentials were employed for carbon, 
where valence states included the $2s$ and $2p$ electrons. The plane-wave basis set cutoff was set to 750 eV, ensuring 
that the total energy and unit cell volume were converged, as demonstrated in the study of crystalline indole and 
tetracyanoethylene~\cite{Bohdan-indole,Bohdan-TCNE}. 
The \textit{k}-point grid was kept to maintain a spacing of 0.07 {\AA}$^{-1}$.  
Explicit all-electron calculations using the FHI-aims code~\cite{aims} confirm that the
binding energies from the pseudopotential calculations are converged to better than 0.01 eV per molecule.

The optimized PBE+vdW lattice parameters for polyacenes along with X-ray measurement results are shown in Table~\ref{tabGeom}.
The overall deviations between our calculations and experiments are less than 2\% in lattice parameters and unit-cell volumes. 
Similar agreement is also found for the internal molecular geometries, where the PBE+vdW method predicts the C--C distances with an 
accuracy of 2\% in comparison with X-ray measurements. The polyacene crystal densities predicted by 
the PBE+vdW method are slightly higher than the experimental ones (except for pentacene), consistent with the fact
that the inclusion of zero-point energy and thermal expansion will decrease the density of the crystal.
\begin{table}
\caption{Unit cell parameters for polyacene molecular crystals determined from PBE+vdW calculations
and low temperature X-ray experiments.
The dielectric constants are reported using the Clausius-Mosotti equation corresponding to
Eq.~\ref{eqPade} ($\varepsilon_h$), and Eq.~\ref{eqScr} ($\varepsilon_{\rm{full}}$). 
Data are reported for naphthalene (2A), anthracene (3A), tetracene (4A), and pentacene (5A).}
\label{tabGeom}
\begin{tabular}{cccccccccc}
\hline\hline
& $a$ ({\AA}) & $b$ ({\AA}) & $c$ ({\AA}) & $\rho$ (g/ml) & $\varepsilon_{\rm{full}}$ & $\varepsilon_h$ \\
\hline
2A   & 8.117 & 5.897 & 8.647 & 1.244 & 3.06 & 3.58 \\
Exp. & 8.108 & 5.940 & 8.647 & 1.239 & 3.2~\cite{naphthalene-dielectric}\\
\hline
3A   & 8.399 & 5.906 & 11.120 & 1.313 & 3.24 & 3.80 \\
Exp. & 8.414 & 5.990 & 11.095 & 1.297 & 3.2~\cite{OrgCrys-book}\\
\hline
4A   & 6.050 & 7.706 & 13.030 & 1.343 & 3.31 & 3.89\\ 
Exp. & 6.056 & 7.838 & 13.010 & 1.323 & -- & \\
\hline
5A   & 6.129 & 7.676 & 14.531 & 1.392 & 3.44 & 4.08\\
Exp. & 6.239 & 7.636 & 14.333 & 1.397 & 2.7--3.89~\cite{pentacene-dielectric-anisotropy,pentacene-dielectric}\\
\hline\hline
\end{tabular}
\end{table}

To illustrate the importance of electrodynamic response in polyacene crystals, we have
computed the solid dielectric constant, $\varepsilon$, using the Clausius-Mossotti formula.  
The required polarizabilities were obtained from: (i) Eq.~(\ref{eqPade}), which only includes local hybridization effects, 
and (ii) Eq.~(\ref{eqScr}), which properly accounts for electrodynamic response screening.
Comparison between the ``hybridized'' $\varepsilon_h$ and the ``full'' $\varepsilon_{\rm{full}}$ 
dielectric constants in Table~\ref{tabGeom} reveals the importance of electrodynamic interactions. 
The $\varepsilon_{\rm{full}}$ dielectric constants for all polyacene solids are decreased by 15\%   
when compared with $\varepsilon_h$. The fully screened dielectric constants are in excellent
agreement with the measured values of $\varepsilon$~\cite{OrgCrys-book,naphthalene-dielectric,pentacene-dielectric-anisotropy,pentacene-dielectric}. 
For pentacene, $\varepsilon_{\rm{full}}$ is also close to the values of 3.2 and 3.6, 
obtained by Sharifzadeh \emph{et al.}~\cite{LK-JF-PRB-2012} from $GW$ calculations within the random phase approximation (RPA). 
Indeed, the SCS method solves the RPA equation for a collection of QHOs in the dipole approximation. 
This explains the good agreement with $GW$ dielectric constant and also the fact that the SCS model successfully 
reproduces the measured dielectric constants of crystalline silicon and germanium~\cite{MBD,Vivek}. We note that
the calculated dielectric constants from SCS allow us to approximately determine the fundamental band-gap of a molecular crystal.
The fundamental gap of a molecular crystal is reduced with respect to that of a molecule in the gas phase, owing to the dielectric screening, 
which reduces the energy needed for adding or removing an electron. To evaluate the fundamental gap of crystalline pentacene we may 
use the polarization model, whereby the gap of the gas phase molecule is reduced by $2P$. The polarization energy, $P$ (in atomic units), 
is given by $-(\varepsilon - 1)/(2 R \varepsilon)$, where $R$ is determined from the volume per molecule 
in the unit cell as: $R = [3V_{\rm{cell}}/(8\pi)]^{1/3}$ (for two molecules per cell)~\cite{Soos-CPL-2000,Neaton-PRL,LK-JF-PRB-2012}. 
Using cell parameters calculated with PBE+vdW and $\varepsilon_{\rm{full}}$ we obtain $P$=--1.17 eV. Applying the polarization model 
to the gas-phase gap of 4.57 eV, obtained from a $GW$ calculation based on a consistent starting point, as described in Ref.~\cite{Noa-PRB-2012},
we obtain a bulk gap of 2.22 eV in good agreement with experiment~\cite{pentacene-exp-gap}  
and with explicit $GW$ calculations for pentacene crystal~\cite{LK-JF-PRB-2012}. 
We note that the optical gap is further reduced with respect to the fundamental gap due to excitonic 
effects~\cite{pentacene-dielectric-anisotropy,pentacene-BSE,LK-JF-PRB-2012}, which are not accounted for by the SCS model.

Experimentally, the stability of molecular crystals is measured in terms of their sublimation enthalpy, i.e.
the energy required to convert a certain amount of molecules from the crystalline phase to the
gas phase. The sublimation process is carried out at a given temperature under constant pressure.
The sublimation temperature is largely determined by the
magnitude of the cohesive forces in the crystal. For polyacene crystals, the sublimation temperature
varies from $\approx$ 300 K for naphtalene to $\approx$ 500 K for pentacene~\cite{PAH-JPCRD}, 
illustrating the increase in crystal stability for larger acene molecules.
There are numerous experiments that measure the sublimation enthalpies of
polyacenes; in Table~\ref{tabEner} we report a range of available values, extrapolated to 0 K~\cite{PAH-JPCRD}. We have only taken
those values that are recommended as reliable after critical revision by the authors of Ref.~\cite{PAH-JPCRD}, 
thus avoiding anomalously small or large sublimation enthalpies. 
Both naphthalene and anthracene crystals have been vigorously
studied, and their sublimation enthalpies are well known with a spread of 0.05 eV and 0.12 eV,
respectively. There are fewer measurements available for tetracene and pentacene, and
for the latter the three available experimental values deviate by 0.55 eV. 

In order to compare theoretical lattice energies to
the measured sublimation enthalpies, the sublimation enthalpies need to be extrapolated to zero temperature
by adding the enthalpy difference $[H_{c}^0(T) - H_{c}^0(0)] - [H_{g}^0(T) - H_{g}^0(0)]$,
where the subscript $c$ refers to the crystal, whereas $g$ refers to the gas phase.
In this work, we have calculated this enthalpy difference by integrating the measured heat capacity
$C_p(T)$ for acene crystals, and the extrapolated gas-phase heat capacity from 
Refs.~\cite{NIST-webbook,naphthalene-solid-thermo,anthracene-solid-thermo,tetracene-solid-thermo}. 
Even at 0 K, the sublimation enthalpy includes zero-point vibrational effects, and these have to be
considered when comparing calculated lattice energies to the experimental enthalpy extrapolated to 0 K.
Here we determined the zero-point energy (ZPE) from phonon calculations using the PBE+vdW method with 
the supercell formalism in CASTEP~\cite{CASTEP}. Special care has been taken to converge the supercell size and plane-wave cutoff when
performing phonon calculations. We estimate that our ZPE calculations are converged to 5 meV/molecule. 
Note that vdW interactions contribute significantly to the ZPE energy and have to
be included in phonon calculations to reach this level of accuracy. 

\begin{table}
\caption{Lattice energies of polyacene crystals including zero-point energy (PBE+MBDh and PBE+MBD
calculations using optimized PBE+vdW geometries from Table~\ref{tabGeom}). 
The range of experimental (``Exp.'') ``lattice energies'' from Ref.~\cite{PAH-JPCRD} 
extrapolated to 0 K. 
Also shown are the $\Delta H$ from room temperature (RT) to 0 K calculated
from an integral over experimental $C_p(T)$ data, and the harmonic zero-point energy
calculated using the PBE+vdW method. All values are in units of eV.
}
\label{tabEner}
\begin{tabular}{cccccccccc}
\hline\hline
& $\Delta H($RT$\rightarrow0)$ & ZPE & PBE+MBDh & PBE+MBD & Exp. \\
\hline
%Experimental numbers include the deltaH in them!! Remove it before
%changing any experimental numbers.
2A & 0.041 & 0.069 & -0.993 & -0.862 & -0.803 to -0.752 \\
3A & 0.034 & 0.078 & -1.433 & -1.206 & -1.148 to -1.024 \\
4A & -0.016& 0.110 & -1.951 & -1.587 & -1.525 to -1.299 \\
5A & -0.032& 0.115 & -2.501 & -2.018 & -2.082 to -1.533 \\
\hline\hline
\end{tabular}
\end{table}
\begin{figure}
\includegraphics[scale=0.52]{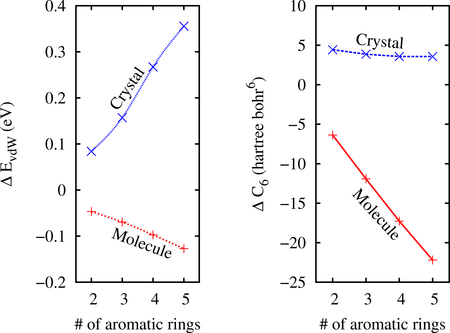}
\caption{(Color online) Left: The difference in the vdW energy between the PBE+MBD
method including electrodynamic response and the PBE+MBDh method based
on ``atom-in-a-molecule'' polarizabilities of Eq.~(\ref{eqPade}).
Right: The difference in the $C_6$ coefficients between PBE+MBD and PBE+MBDh methods per carbon atom.
}\label{figMolCrys}
\end{figure}
Now we analyze the impact of electrodynamic response on the sublimation enthalpies of polyacene crystals. 
For this purpose we combine the PBE functional with the recently developed many-body dispersion (MBD) method~\cite{MBD}. 
The MBD energy expression computes the long-range many-body dispersion energy to infinite order for molecules and solids with a finite
band gap. In the MBD method, the full electronic system is mapped to a system of quantum harmonic oscillators.
The QHO polarizabilities can be obtained either from Eq.~(\ref{eqPade}) (from now on called the DFT+MBDh method) 
or Eq.~(\ref{eqScr}) (from now on called the DFT+MBD method), thus allowing us to clearly assess the effect of 
electrodynamic screening on the dispersion energy and stability of molecular crystals. The ZPE-inclusive lattice energies 
obtained with both methods are shown in Table~\ref{tabEner} and compared with experimental data.

The inspection of Table~\ref{tabEner} illustrates the crucial importance of electrodynamic response
for the stability of polyacene crystals. The PBE+MBDh method uses semilocal hybridized polarizabilities
and overestimates the experimental sublimation enthalpies by more than 0.20 eV for naphthalene and
up to 0.44 eV for pentacene. Upon including the response screening, as depolarization reduces the stability of naphthalene by 0.13 eV 
and of pentacene by 0.48 eV compared to the PBE+MBDh approach. 
To assess the influence of the underlying DFT functional, we have also carried out calculations using the PBE-based hybrid
functional, PBE0~\cite{PBE0-1,PBE0-2}, combined with the MBD method. 
The PBE0 functional describes the electrostatic and inductive intermolecular
interactions more accurately~\cite{MBD,ice-PRL}. However, we found that the PBE0+MBD approach 
yields essentially the same lattice energies as the PBE+MBD method
for all polyacenes (within 10 meV per molecule). 

The remaining slight overestimation of lattice energies in Table~\ref{tabEner} compared to the 
experimental range can be explained by the fact that the sublimation enthalpy is measured at finite temperature, where
the unit cell undergoes thermal expansion. When using the experimental unit cell at 295 K for naphthalene, 
the PBE+MBD method yields a lattice energy that is increased by 50 meV, essentially within the experimental
range reported in Table~\ref{tabEner}. 

Finally, we explain the observed difference between the PBE+MBDh and PBE+MBD methods
by analyzing the vdW dispersion energies in the gas and crystal phases in Figure~\ref{figMolCrys}.
Along with vdW energies, we also show the \emph{change} in the vdW $C_6$ coefficient per carbon atom upon inclusion of the 
electrodynamic response in the MBD method. 
For the gas phase molecules, there is a significant dipole polarization along the long molecular axis, 
which increases the molecular polarizabilities and $C_6$ coefficients, leading to an increase in the 
vdW energy in the PBE+MBD method when compared to PBE+MBDh. 
However, since the vdW energy contribution to the molecular stability is relatively small, 
the change due to electrodynamic response is only --0.05 eV for naphthalene and up to --0.13 eV for
pentacene. The electrodynamic response gives rise to a radically different situation
in the crystal phase; in this case the interaction with neighboring molecules leads
to overall depolarization, \textit{decreasing} the $C_6$ coefficients by roughly a 
constant amount, when compared to the gas-phase molecules.  
However, the vdW energy makes a larger absolute contribution to the stability of acene crystals. 
This explains the sharp decrease of the vdW energy 
in the crystal predicted by the PBE+MBD method when compared to the PBE+MBDh approach. 
Overall, the opposite effect of screening for the gas-phase molecule and the crystal 
explains the significant reduction of the lattice energy, shown in Table~\ref{tabEner}, upon including
electrodynamic response in the PBE+MBD method.

In summary, we have quantitatively established a connection between collective electrodynamic response 
and the stability of molecular crystals.  
Our results demonstrate that molecular crystals are significantly more complex than a simple collection
of constituent molecules, and that electrodynamic response is crucial
for explaining many of the unique properties of molecular crystals.
We provide to our knowledge the first quantification of the influence of electrodynamic
response on sublimation enthalpies of polyacene crystals, and it is not unreasonable to expect 
that our findings will hold in general for other classes of molecular solids.

A.T. acknowledges support from the European Research Council (ERC Starting Grant \texttt{VDW-CMAT}).
B.S. acknowledges support from the Eberly Science Foundation.

\bibliography{literature}

%Merlin.mbs v4.21 2009-07-09.
\begin{thebibliography}{10}%
\makeatletter
\providecommand \@ifxundefined [1]{%
 \ifx #1\undefined \expandafter \@firstoftwo
 \else \expandafter \@secondoftwo
\fi
}%
\providecommand \@ifnum [1]{%
 \ifnum #1\expandafter \@firstoftwo
 \else \expandafter \@secondoftwo
\fi
}%
\providecommand \enquote [1]{``#1''}%
\providecommand \bibnamefont  [1]{#1}%
\providecommand \bibfnamefont [1]{#1}%
\providecommand \citenamefont [1]{#1}%
\providecommand\href[0]{\@sanitize\@href}%
\providecommand\@href[1]{\endgroup\@@startlink{#1}\endgroup\@@href}%
\providecommand\@@href[1]{#1\@@endlink}%
\providecommand \@sanitize [0]{\begingroup\catcode`\&12\catcode`\#12\relax}%
\@ifxundefined \pdfoutput {\@firstoftwo}{%
 \@ifnum{\z@=\pdfoutput}{\@firstoftwo}{\@secondoftwo}%
}{%
 \providecommand\@@startlink[1]{\leavevmode\special{html:<a href="#1">}}%
 \providecommand\@@endlink[0]{\special{html:</a>}}%
}{%
 \providecommand\@@startlink[1]{%
  \leavevmode
  \pdfstartlink
   attr{/Border[0 0 1 ]/H/I/C[0 1 1]}%
   user{/Subtype/Link/A<</Type/Action/S/URI/URI(#1)>>}%
  \relax
 }%
 \providecommand\@@endlink[0]{\pdfendlink}%
}%
\providecommand \url  [0]{\begingroup\@sanitize \@url }%
\providecommand \@url [1]{\endgroup\@href {#1}{\urlprefix}}%
\providecommand \urlprefix [0]{URL }%
\providecommand \Eprint[0]{\href }%
\@ifxundefined \urlstyle {%
  \providecommand \doi [1]{doi:\discretionary{}{}{}#1}%
}{%
  \providecommand \doi [0]{doi:\discretionary{}{}{}\begingroup
  \urlstyle{rm}\Url }%
}%
\providecommand \doibase [0]{http://dx.doi.org/}%
\providecommand \Doi[1]{\href{\doibase#1}}%
\providecommand \bibAnnote [3]{%
  \BibitemShut{#1}%
  \begin{quotation}\noindent
    \textsc{Key:}\ #2\\\textsc{Annotation:}\ #3%
  \end{quotation}%
}%
\providecommand \bibAnnoteFile [2]{%
  \IfFileExists{#2}{\bibAnnote {#1} {#2} {\input{#2}}}{}%
}%
\providecommand \typeout [0]{\immediate \write \m@ne }%
\providecommand \selectlanguage [0]{\@gobble}%
\providecommand \bibinfo [0]{\@secondoftwo}%
\providecommand \bibfield [0]{\@secondoftwo}%
\providecommand \translation [1]{[#1]}%
\providecommand \BibitemOpen[0]{}%
\providecommand \bibitemStop [0]{}%
\providecommand \bibitemNoStop [0]{.\EOS\space}%
\providecommand \EOS [0]{\spacefactor3000\relax}%
\providecommand \BibitemShut [1]{\csname bibitem#1\endcsname}%
%</preamble>
\bibitem{Forrest-Nature}%
  \BibitemOpen
  \bibfield{author}{%
  \bibinfo {author} {\bibfnamefont{S.~R.}\ \bibnamefont{Forrest}},\ }%
  \bibfield{journal}{%
  \bibinfo {journal} {Nature}\ }%
  \textbf{\bibinfo {volume} {428}},\ \bibinfo {pages} {911} (\bibinfo {year}
  {2004})%
  \bibAnnoteFile{NoStop}{Forrest-Nature}%
\bibitem{ChemMater-review}%
  \BibitemOpen
  \bibfield{author}{%
  \bibinfo {author} {\bibfnamefont{T.~W.}\ \bibnamefont{Kelley}}, \bibinfo
  {author} {\bibfnamefont{P.~F.}\ \bibnamefont{Baude}}, \bibinfo {author}
  {\bibfnamefont{C.}~\bibnamefont{Gerlach}}, \bibinfo {author}
  {\bibfnamefont{D.~E.}\ \bibnamefont{Ender}}, \bibinfo {author}
  {\bibfnamefont{D.}~\bibnamefont{Muyres}}, \bibinfo {author}
  {\bibfnamefont{M.~A.}\ \bibnamefont{Haase}}, \bibinfo {author}
  {\bibfnamefont{D.~E.}\ \bibnamefont{Vogel}},\ and\ \bibinfo {author}
  {\bibfnamefont{S.~D.}\ \bibnamefont{Theiss}},\ }%
  \bibfield{journal}{%
  \bibinfo {journal} {Chem. Mater.}\ }%
  \textbf{\bibinfo {volume} {16}},\ \bibinfo {pages} {4413} (\bibinfo {year}
  {2004})%
  \bibAnnoteFile{NoStop}{ChemMater-review}%
\bibitem{LK-JF-PRB-2012}%
  \BibitemOpen
  \bibfield{author}{%
  \bibinfo {author} {\bibfnamefont{S.}~\bibnamefont{Sharifzadeh}}, \bibinfo
  {author} {\bibfnamefont{A.}~\bibnamefont{Biller}}, \bibinfo {author}
  {\bibfnamefont{L.}~\bibnamefont{Kronik}},\ and\ \bibinfo {author}
  {\bibfnamefont{J.~B.}\ \bibnamefont{Neaton}},\ }%
  \bibfield{journal}{%
  \bibinfo {journal} {Phys. Rev. B}\ }%
  \textbf{\bibinfo {volume} {85}},\ \bibinfo {pages} {125307} (\bibinfo {year}
  {2012})%
  \bibAnnoteFile{NoStop}{LK-JF-PRB-2012}%
\bibitem{OrgCrys-book}%
  \BibitemOpen
  \bibfield{author}{%
  \bibinfo {author} {\bibfnamefont{M.}~\bibnamefont{Pope}}\ and\ \bibinfo
  {author} {\bibfnamefont{C.~E.}\ \bibnamefont{Swenberg}},\ }%
  \emph{\bibinfo {title} {Electronic Processes in Organic Crystals and
  Polymers}}\ (\bibinfo {publisher} {Oxford University Press, New York},\
  \bibinfo {year} {1999})%
  \bibAnnoteFile{NoStop}{OrgCrys-book}%
\bibitem{JEAnthony-Angewandte-2007}%
  \BibitemOpen
  \bibfield{author}{%
  \bibinfo {author} {\bibfnamefont{J.~E.}\ \bibnamefont{Anthony}},\ }%
  \bibfield{journal}{%
  \bibinfo {journal} {Angew. Chem. Int. Ed.}\ }%
  \textbf{\bibinfo {volume} {47}},\ \bibinfo {pages} {452} (\bibinfo {year}
  {2008})%
  \bibAnnoteFile{NoStop}{JEAnthony-Angewandte-2007}%
\bibitem{Toll-PR-1956}%
  \BibitemOpen
  \bibfield{author}{%
  \bibinfo {author} {\bibfnamefont{J.~S.}\ \bibnamefont{Toll}},\ }%
  \bibfield{journal}{%
  \bibinfo {journal} {Phys. Rev. B}\ }%
  \textbf{\bibinfo {volume} {104}},\ \bibinfo {pages} {1760} (\bibinfo {year}
  {1956})%
  \bibAnnoteFile{NoStop}{Toll-PR-1956}%
\bibitem{Neumann}%
  \BibitemOpen
  \bibfield{author}{%
  \bibinfo {author} {\bibfnamefont{M.~A.}\ \bibnamefont{Neumann}}\ and\
  \bibinfo {author} {\bibfnamefont{M.~A.}\ \bibnamefont{Perrin}},\ }%
  \bibfield{journal}{%
  \bibinfo {journal} {J. Phys. Chem. B}\ }%
  \textbf{\bibinfo {volume} {109}},\ \bibinfo {pages} {15531} (\bibinfo {year}
  {2005})%
  \bibAnnoteFile{NoStop}{Neumann}%
\bibitem{CSP2010}%
  \BibitemOpen
  \bibfield{author}{%
  \bibinfo {author} {\bibnamefont{{D. A. Bardwell \textit{et al.}}}},\ }%
  \bibfield{journal}{%
  \bibinfo {journal} {Acta. Cryst. B}\ }%
  \textbf{\bibinfo {volume} {67}},\ \bibinfo {pages} {535} (\bibinfo {year}
  {2011})%
  \bibAnnoteFile{NoStop}{CSP2010}%
\bibitem{Beran-review}%
  \BibitemOpen
  \bibfield{author}{%
  \bibinfo {author} {\bibfnamefont{S.}~\bibnamefont{Wen}}, \bibinfo {author}
  {\bibfnamefont{K.}~\bibnamefont{Nanda}}, \bibinfo {author}
  {\bibfnamefont{Y.}~\bibnamefont{Huang}},\ and\ \bibinfo {author}
  {\bibfnamefont{G.}~\bibnamefont{Beran}},\ }%
  \bibfield{journal}{%
  \bibinfo {journal} {Phys. Chem. Chem. Phys.}\ }%
  \textbf{\bibinfo {volume} {14}},\ \bibinfo {pages} {7578} (\bibinfo {year}
  {2012})%
  \bibAnnoteFile{NoStop}{Beran-review}%
\bibitem{Stone-book}%
  \BibitemOpen
  \bibfield{author}{%
  \bibinfo {author} {\bibfnamefont{A.~J.}\ \bibnamefont{Stone}},\ }%
  \emph{\bibinfo {title} {The theory of intermolecular forces}}\ (\bibinfo
  {publisher} {Oxford University Press},\ \bibinfo {year} {1996})%
  \bibAnnoteFile{NoStop}{Stone-book}%
\bibitem{TS-vdW}%
  \BibitemOpen
  \bibfield{author}{%
  \bibinfo {author} {\bibfnamefont{A.}~\bibnamefont{Tkatchenko}}\ and\ \bibinfo
  {author} {\bibfnamefont{M.}~\bibnamefont{Scheffler}},\ }%
  \bibfield{journal}{%
  \bibinfo {journal} {Phys. Rev. Lett.}\ }%
  \textbf{\bibinfo {volume} {102}},\ \bibinfo {pages} {073005} (\bibinfo {year}
  {2009})%
  \bibAnnoteFile{NoStop}{TS-vdW}%
\bibitem{pentacene-BSE}%
  \BibitemOpen
  \bibfield{author}{%
  \bibinfo {author} {\bibfnamefont{M.~L.}\ \bibnamefont{Tiago}}, \bibinfo
  {author} {\bibfnamefont{J.~E.}\ \bibnamefont{Northrup}},\ and\ \bibinfo
  {author} {\bibfnamefont{S.~G.}\ \bibnamefont{Louie}},\ }%
  \bibfield{journal}{%
  \bibinfo {journal} {Phys. Rev. B}\ }%
  \textbf{\bibinfo {volume} {67}},\ \bibinfo {pages} {115212} (\bibinfo {year}
  {2003})%
  \bibAnnoteFile{NoStop}{pentacene-BSE}%
\bibitem{MBD}%
  \BibitemOpen
  \bibfield{author}{%
  \bibinfo {author} {\bibfnamefont{A.}~\bibnamefont{Tkatchenko}}, \bibinfo
  {author} {\bibnamefont{{R. A. DiStasio, Jr.}}}, \bibinfo {author}
  {\bibfnamefont{R.}~\bibnamefont{Car}},\ and\ \bibinfo {author}
  {\bibfnamefont{M.}~\bibnamefont{Scheffler}},\ }%
  \bibfield{journal}{%
  \bibinfo {journal} {Phys. Rev. Lett.}\ }%
  \textbf{\bibinfo {volume} {108}},\ \bibinfo {pages} {236402} (\bibinfo {year}
  {2012})%
  \bibAnnoteFile{NoStop}{MBD}%
\bibitem{Felderhof}%
  \BibitemOpen
  \bibfield{author}{%
  \bibinfo {author} {\bibfnamefont{B.~U.}\ \bibnamefont{Felderhof}},\ }%
  \bibfield{journal}{%
  \bibinfo {journal} {Physica}\ }%
  \textbf{\bibinfo {volume} {76}},\ \bibinfo {pages} {486} (\bibinfo {year}
  {1974})%
  \bibAnnoteFile{NoStop}{Felderhof}%
\bibitem{Oxtoby-Gelbart}%
  \BibitemOpen
  \bibfield{author}{%
  \bibinfo {author} {\bibfnamefont{D.~W.}\ \bibnamefont{Oxtoby}}\ and\ \bibinfo
  {author} {\bibfnamefont{W.~M.}\ \bibnamefont{Gelbart}},\ }%
  \bibfield{journal}{%
  \bibinfo {journal} {Mol. Phys.}\ }%
  \textbf{\bibinfo {volume} {29}},\ \bibinfo {pages} {1569} (\bibinfo {year}
  {1975})%
  \bibAnnoteFile{NoStop}{Oxtoby-Gelbart}%
\bibitem{Thole}%
  \BibitemOpen
  \bibfield{author}{%
  \bibinfo {author} {\bibfnamefont{B.~T.}\ \bibnamefont{Thole}},\ }%
  \bibfield{journal}{%
  \bibinfo {journal} {Chem. Phys.}\ }%
  \textbf{\bibinfo {volume} {59}},\ \bibinfo {pages} {341} (\bibinfo {year}
  {1981})%
  \bibAnnoteFile{NoStop}{Thole}%
\bibitem{pentacene-dielectric-anisotropy}%
  \BibitemOpen
  \bibfield{author}{%
  \bibinfo {author} {\bibfnamefont{D.}~\bibnamefont{Faltermeier}}, \bibinfo
  {author} {\bibfnamefont{B.}~\bibnamefont{Gompf}}, \bibinfo {author}
  {\bibfnamefont{M.}~\bibnamefont{Dressel}}, \bibinfo {author}
  {\bibfnamefont{A.~K.}\ \bibnamefont{Tripathi}},\ and\ \bibinfo {author}
  {\bibfnamefont{J.}~\bibnamefont{Pflaum}},\ }%
  \bibfield{journal}{%
  \bibinfo {journal} {Phys. Rev. B}\ }%
  \textbf{\bibinfo {volume} {74}},\ \bibinfo {pages} {125416} (\bibinfo {year}
  {2006})%
  \bibAnnoteFile{NoStop}{pentacene-dielectric-anisotropy}%
\bibitem{CSD}%
  \BibitemOpen
  \bibfield{author}{%
  \bibinfo {author} {\bibfnamefont{F.~H.}\ \bibnamefont{Allen}},\ }%
  \bibfield{journal}{%
  \bibinfo {journal} {Acta. Crys.}\ }%
  \textbf{\bibinfo {volume} {B58}},\ \bibinfo {pages} {380} (\bibinfo {year}
  {2002})%
  \bibAnnoteFile{NoStop}{CSD}%
\bibitem{PBE}%
  \BibitemOpen
  \bibfield{author}{%
  \bibinfo {author} {\bibfnamefont{J.~P.}\ \bibnamefont{Perdew}}, \bibinfo
  {author} {\bibfnamefont{K.}~\bibnamefont{Burke}},\ and\ \bibinfo {author}
  {\bibfnamefont{M.}~\bibnamefont{Ernzerhof}},\ }%
  \bibfield{journal}{%
  \bibinfo {journal} {Phys. Rev. Lett.}\ }%
  \textbf{\bibinfo {volume} {77}},\ \bibinfo {pages} {3865} (\bibinfo {year}
  {1996})%
  \bibAnnoteFile{NoStop}{PBE}%
\bibitem{CASTEP}%
  \BibitemOpen
  \bibfield{author}{%
  \bibinfo {author} {\bibfnamefont{S.~J.}\ \bibnamefont{Clark}}, \bibinfo
  {author} {\bibfnamefont{M.~D.}\ \bibnamefont{Segall}}, \bibinfo {author}
  {\bibfnamefont{C.~J.}\ \bibnamefont{Pickard}}, \bibinfo {author}
  {\bibfnamefont{P.~J.}\ \bibnamefont{Hasnip}}, \bibinfo {author}
  {\bibfnamefont{M.~J.}\ \bibnamefont{Probert}}, \bibinfo {author}
  {\bibfnamefont{K.}~\bibnamefont{Refson}},\ and\ \bibinfo {author}
  {\bibfnamefont{M.~C.}\ \bibnamefont{Payne}},\ }%
  \bibfield{journal}{%
  \bibinfo {journal} {Zeit. Krist.}\ }%
  \textbf{\bibinfo {volume} {220}},\ \bibinfo {pages} {567} (\bibinfo {year}
  {2005})%
  \bibAnnoteFile{NoStop}{CASTEP}%
\bibitem{Erik}%
  \BibitemOpen
  \bibfield{author}{%
  \bibinfo {author} {\bibfnamefont{E.~R.}\ \bibnamefont{McNellis}}, \bibinfo
  {author} {\bibfnamefont{J.}~\bibnamefont{Meyer}},\ and\ \bibinfo {author}
  {\bibfnamefont{K.}~\bibnamefont{Reuter}},\ }%
  \bibfield{journal}{%
  \bibinfo {journal} {Phys. Rev. B}\ }%
  \textbf{\bibinfo {volume} {80}},\ \bibinfo {pages} {205414} (\bibinfo {year}
  {2009})%
  \bibAnnoteFile{NoStop}{Erik}%
\bibitem{Bohdan-indole}%
  \BibitemOpen
  \bibfield{author}{%
  \bibinfo {author} {\bibfnamefont{B.}~\bibnamefont{Schatschneider}}\ and\
  \bibinfo {author} {\bibfnamefont{J.-J.}\ \bibnamefont{Liang}},\ }%
  \bibfield{journal}{%
  \bibinfo {journal} {J. Chem. Phys.}\ }%
  \textbf{\bibinfo {volume} {135}},\ \bibinfo {pages} {164508} (\bibinfo {year}
  {2011})%
  \bibAnnoteFile{NoStop}{Bohdan-indole}%
\bibitem{Bohdan-TCNE}%
  \BibitemOpen
  \bibfield{author}{%
  \bibinfo {author} {\bibfnamefont{B.}~\bibnamefont{Schatschneider}}, \bibinfo
  {author} {\bibfnamefont{J.-J.}\ \bibnamefont{Liang}}, \bibinfo {author}
  {\bibfnamefont{S.}~\bibnamefont{Jezowksi}},\ and\ \bibinfo {author}
  {\bibfnamefont{A.}~\bibnamefont{Tkatchenko}},\ }%
  \bibfield{journal}{%
  \bibinfo {journal} {CrystEngComm}\ }%
  \textbf{\bibinfo {volume} {14}},\ \bibinfo {pages} {4656} (\bibinfo {year}
  {2012})%
  \bibAnnoteFile{NoStop}{Bohdan-TCNE}%
\bibitem{aims}%
  \BibitemOpen
  \bibfield{author}{%
  \bibinfo {author} {\bibfnamefont{V.}~\bibnamefont{Blum}}, \bibinfo {author}
  {\bibfnamefont{R.}~\bibnamefont{Gehrke}}, \bibinfo {author}
  {\bibfnamefont{F.}~\bibnamefont{Hanke}}, \bibinfo {author}
  {\bibfnamefont{P.}~\bibnamefont{Havu}}, \bibinfo {author}
  {\bibfnamefont{V.}~\bibnamefont{Havu}}, \bibinfo {author}
  {\bibfnamefont{X.}~\bibnamefont{Ren}}, \bibinfo {author}
  {\bibfnamefont{K.}~\bibnamefont{Reuter}},\ and\ \bibinfo {author}
  {\bibfnamefont{M.}~\bibnamefont{Scheffler}},\ }%
  \bibfield{journal}{%
  \bibinfo {journal} {Comp. Phys. Comm.}\ }%
  \textbf{\bibinfo {volume} {180}},\ \bibinfo {pages} {2175} (\bibinfo {year}
  {2009})%
  \bibAnnoteFile{NoStop}{aims}%
\bibitem{naphthalene-dielectric}%
  \BibitemOpen
  \bibfield{author}{%
  \bibinfo {author} {\bibfnamefont{T.}~\bibnamefont{Suthana}}, \bibinfo
  {author} {\bibfnamefont{N.}~\bibnamefont{Rajesha}}, \bibinfo {author}
  {\bibfnamefont{P.}~\bibnamefont{Dhanaraja}},\ and\ \bibinfo {author}
  {\bibfnamefont{C.}~\bibnamefont{Mahadevan}},\ }%
  \bibfield{journal}{%
  \bibinfo {journal} {Spectrochimica Acta Part A}\ }%
  \textbf{\bibinfo {volume} {75}},\ \bibinfo {pages} {69} (\bibinfo {year}
  {2010})%
  \bibAnnoteFile{NoStop}{naphthalene-dielectric}%
\bibitem{pentacene-dielectric}%
  \BibitemOpen
  \bibfield{author}{%
  \bibinfo {author} {\bibfnamefont{C.~H.}\ \bibnamefont{Kim}}, \bibinfo
  {author} {\bibfnamefont{O.}~\bibnamefont{Yaghmazadeh}}, \bibinfo {author}
  {\bibfnamefont{D.}~\bibnamefont{Tondelier}}, \bibinfo {author}
  {\bibfnamefont{Y.~B.}\ \bibnamefont{Jeong}}, \bibinfo {author}
  {\bibfnamefont{Y.}~\bibnamefont{Bonnassieux}},\ and\ \bibinfo {author}
  {\bibfnamefont{G.}~\bibnamefont{Horowitz}},\ }%
  \bibfield{journal}{%
  \bibinfo {journal} {J. Appl. Phys.}\ }%
  \textbf{\bibinfo {volume} {109}},\ \bibinfo {pages} {083710} (\bibinfo {year}
  {2011})%
  \bibAnnoteFile{NoStop}{pentacene-dielectric}%
\bibitem{Vivek}%
  \BibitemOpen
  \bibinfo {note} {{V. V. Gobre \textit{et al.}, to be published}}%
  \bibAnnoteFile{NoStop}{Vivek}%
\bibitem{Soos-CPL-2000}%
  \BibitemOpen
  \bibfield{author}{%
  \bibinfo {author} {\bibfnamefont{I.~G.}\ \bibnamefont{Hill}}, \bibinfo
  {author} {\bibfnamefont{A.}~\bibnamefont{Kahn}}, \bibinfo {author}
  {\bibfnamefont{Z.~G.}\ \bibnamefont{Soos}},\ and\ \bibinfo {author}
  {\bibnamefont{{R. A. Pascal, Jr.}}}\ }%
  \textbf{\bibinfo {volume} {327}},\ \bibinfo {pages} {181} (\bibinfo {year}
  {2000})%
  \bibAnnoteFile{NoStop}{Soos-CPL-2000}%
\bibitem{Neaton-PRL}%
  \BibitemOpen
  \bibfield{author}{%
  \bibinfo {author} {\bibfnamefont{J.~B.}\ \bibnamefont{Neaton}}, \bibinfo
  {author} {\bibfnamefont{M.~S.}\ \bibnamefont{Hybertsen}},\ and\ \bibinfo
  {author} {\bibfnamefont{S.~G.}\ \bibnamefont{Louie}},\ }%
  \bibfield{journal}{%
  \bibinfo {journal} {Phys. Rev. Lett.}\ }%
  \textbf{\bibinfo {volume} {97}},\ \bibinfo {pages} {216405} (\bibinfo {year}
  {2006})%
  \bibAnnoteFile{NoStop}{Neaton-PRL}%
\bibitem{Noa-PRB-2012}%
  \BibitemOpen
  \bibfield{author}{%
  \bibinfo {author} {\bibfnamefont{T.}~\bibnamefont{K\"orzd\"orfer}}\ and\
  \bibinfo {author} {\bibfnamefont{N.}~\bibnamefont{Marom}},\ }%
  \bibfield{journal}{%
  \bibinfo {journal} {Phys. Rev. B}\ }%
  \textbf{\bibinfo {volume} {86}},\ \bibinfo {pages} {041110} (\bibinfo {year}
  {2012})%
  \bibAnnoteFile{NoStop}{Noa-PRB-2012}%
\bibitem{pentacene-exp-gap}%
  \BibitemOpen
  \bibfield{author}{%
  \bibinfo {author} {\bibfnamefont{E.~A.}\ \bibnamefont{Silinsh}}, \bibinfo
  {author} {\bibfnamefont{V.~A.}\ \bibnamefont{Kolesnikov}}, \bibinfo {author}
  {\bibfnamefont{I.~J.}\ \bibnamefont{Muzikante}},\ and\ \bibinfo {author}
  {\bibfnamefont{D.~R.}\ \bibnamefont{Balode}},\ }%
  \bibfield{journal}{%
  \bibinfo {journal} {physica status solidi (b)}\ }%
  \textbf{\bibinfo {volume} {113}},\ \bibinfo {pages} {379} (\bibinfo {year}
  {1982})%
  \bibAnnoteFile{NoStop}{pentacene-exp-gap}%
\bibitem{PAH-JPCRD}%
  \BibitemOpen
  \bibfield{author}{%
  \bibinfo {author} {\bibfnamefont{M.~V.}\ \bibnamefont{Roux}}, \bibinfo
  {author} {\bibfnamefont{M.}~\bibnamefont{Temprado}}, \bibinfo {author}
  {\bibfnamefont{J.~S.}\ \bibnamefont{Chickos}},\ and\ \bibinfo {author}
  {\bibfnamefont{Y.}~\bibnamefont{Nagano}},\ }%
  \bibfield{journal}{%
  \bibinfo {journal} {J. Phys. Chem. Ref. Data}\ }%
  \textbf{\bibinfo {volume} {37}},\ \bibinfo {pages} {1855} (\bibinfo {year}
  {2008})%
  \bibAnnoteFile{NoStop}{PAH-JPCRD}%
\bibitem{NIST-webbook}%
  \BibitemOpen
  \bibinfo {note} {{NIST WebBook and references within,
  http://webbook.nist.gov/chemistry/name-ser.html.}}%
  \bibAnnoteFile{Stop}{NIST-webbook}%
\bibitem{naphthalene-solid-thermo}%
  \BibitemOpen
  \bibfield{author}{%
  \bibinfo {author} {\bibfnamefont{J.~P.}\ \bibnamefont{McCullough}}, \bibinfo
  {author} {\bibfnamefont{H.~L.}\ \bibnamefont{Finke}}, \bibinfo {author}
  {\bibfnamefont{J.~F.}\ \bibnamefont{Messerly}}, \bibinfo {author}
  {\bibfnamefont{T.~C.}\ \bibnamefont{Kincheloe}},\ and\ \bibinfo {author}
  {\bibfnamefont{G.}~\bibnamefont{Waddington}},\ }%
  \bibfield{journal}{%
  \bibinfo {journal} {J. Phys. Chem.}\ }%
  \textbf{\bibinfo {volume} {61}},\ \bibinfo {pages} {1105} (\bibinfo {year}
  {1957})%
  \bibAnnoteFile{NoStop}{naphthalene-solid-thermo}%
\bibitem{anthracene-solid-thermo}%
  \BibitemOpen
  \bibfield{author}{%
  \bibinfo {author} {\bibfnamefont{P.}~\bibnamefont{Goursot}}, \bibinfo
  {author} {\bibfnamefont{H.~L.}\ \bibnamefont{Girdhar}},\ and\ \bibinfo
  {author} {\bibnamefont{{E. F. Westrum, Jr.}}},\ }%
  \bibfield{journal}{%
  \bibinfo {journal} {J. Phys. Chem.}\ }%
  \textbf{\bibinfo {volume} {74}},\ \bibinfo {pages} {2538} (\bibinfo {year}
  {1970})%
  \bibAnnoteFile{NoStop}{anthracene-solid-thermo}%
\bibitem{tetracene-solid-thermo}%
  \BibitemOpen
  \bibfield{author}{%
  \bibinfo {author} {\bibfnamefont{W.-K.}\ \bibnamefont{Wong}}\ and\ \bibinfo
  {author} {\bibnamefont{{E. F. Westrum, Jr.}}},\ }%
  \bibfield{journal}{%
  \bibinfo {journal} {Mol. Crys. Liq. Cryst.}\ }%
  \textbf{\bibinfo {volume} {61}},\ \bibinfo {pages} {207} (\bibinfo {year}
  {1980})%
  \bibAnnoteFile{NoStop}{tetracene-solid-thermo}%
\bibitem{PBE0-1}%
  \BibitemOpen
  \bibfield{author}{%
  \bibinfo {author} {\bibfnamefont{J.~P.}\ \bibnamefont{Perdew}}, \bibinfo
  {author} {\bibfnamefont{M.}~\bibnamefont{Ernzerhof}},\ and\ \bibinfo {author}
  {\bibfnamefont{K.}~\bibnamefont{Burke}},\ }%
  \bibfield{journal}{%
  \bibinfo {journal} {J. Comp. Phys.}\ }%
  \textbf{\bibinfo {volume} {105}},\ \bibinfo {pages} {9982} (\bibinfo {year}
  {1996})%
  \bibAnnoteFile{NoStop}{PBE0-1}%
\bibitem{PBE0-2}%
  \BibitemOpen
  \bibfield{author}{%
  \bibinfo {author} {\bibfnamefont{C.}~\bibnamefont{Adamo}}\ and\ \bibinfo
  {author} {\bibfnamefont{V.}~\bibnamefont{Barone}},\ }%
  \bibfield{journal}{%
  \bibinfo {journal} {J. Comp. Phys.}\ }%
  \textbf{\bibinfo {volume} {110}},\ \bibinfo {pages} {6158} (\bibinfo {year}
  {1999})%
  \bibAnnoteFile{NoStop}{PBE0-2}%
\bibitem{ice-PRL}%
  \BibitemOpen
  \bibfield{author}{%
  \bibinfo {author} {\bibfnamefont{B.}~\bibnamefont{Santra}}, \bibinfo {author}
  {\bibnamefont{{J. Klime{\v{s}}}}}, \bibinfo {author}
  {\bibfnamefont{D.}~\bibnamefont{Alf{\`e}}}, \bibinfo {author}
  {\bibfnamefont{A.}~\bibnamefont{Tkatchenko}}, \bibinfo {author}
  {\bibfnamefont{B.}~\bibnamefont{Slater}}, \bibinfo {author}
  {\bibfnamefont{A.}~\bibnamefont{Michaelides}}, \bibinfo {author}
  {\bibfnamefont{R.}~\bibnamefont{Car}},\ and\ \bibinfo {author}
  {\bibfnamefont{M.}~\bibnamefont{Scheffler}},\ }%
  \bibfield{journal}{%
  \bibinfo {journal} {Phys. Rev. Lett.}\ }%
  \textbf{\bibinfo {volume} {107}},\ \bibinfo {pages} {185701} (\bibinfo {year}
  {2011})%
  \bibAnnoteFile{NoStop}{ice-PRL}%
\end{thebibliography}%
\end{document}